\documentclass{llncs}
\usepackage{times}

\usepackage{pstricks}
\usepackage{pst-node}
\usepackage{pstcolors}

\usepackage{graphicx} 

\usepackage{hyperref}
\hypersetup{backref,
colorlinks=true,
linkcolor=blue,
urlcolor=blue}

\usepackage{amsfonts}
\usepackage{amssymb}

\usepackage[fleqn,centertags]{amsmath}

\usepackage{theorem}
\theoremheaderfont{\scshape}
\theorembodyfont{\slshape}

\usepackage{stackengine}

\usepackage{abbrev}

\newcommand{\couple}[2]{\ensuremath\langle#1,#2\rangle}

\newcommand{\CAL}[1]{\ensuremath{\mathcal{#1}}}
\newcommand{\T}{\CAL{T}}
\newcommand{\V}{\CAL{V}}

\newcommand{\arrow}{\ensuremath{\rightarrow}}

\newcommand{\eqd}{\ensuremath{\;\stackrel{\mbox{\tiny def}}{=}\;}}

\renewcommand{\empty}{\ensuremath{\emptyset}}

\newcommand{\flink}[1]{\footnote{\href{#1}{\scriptsize\texttt{#1}}}}

\newcommand{\online}[1]{[\href{#1}{online}]}
\newcommand{\ONLINE}[2]{[\href{#1}{online}]}

\input{main.mac}

\input{rules.mac}

\newcommand{\leftfot}{\ensuremath\textbf{t}_1}
\newcommand{\rightfot}{\ensuremath\textbf{t}_2}

\newrgbcolor{nodecolor}{.98 .98 .94}
\newcommand{\TRUE}{\textsc{\textbf{true}}}

\newcommand{\arity}{\textbf{arity}}
\newcommand{\var}{\textbf{var}}
\newcommand{\NN}{\ensuremath\mathrm{I\!N}}

\begin{document}
\title{Lattice Operations on Terms over Similar Signatures}
\author{Hassan A{\"\i}t-Kaci\inst{1} \and
  Gabriella Pasi\inst{2}}
\authorrunning{\textsc{H.~A{\"\i}t-Kaci}; \textsc{G.~Pasi}}
\institute{HAK Language Technologies --- \email{hak@acm.org} \and
  Universit\'a de Milano-Bicocca --- \email{pasi@disco.unimib.it}}
\maketitle
\begin{abstract}
Unification and generalization are operations on two terms computing
respectively their greatest lower bound and least upper bound when the
terms are quasi-ordered by subsumption up to variable renaming (\ie,
$t_1\issubsumedby t_2$ iff $t_1=t_2\sigma$ for some variable
substitution $\sigma$). When term signatures are such that distinct
functor symbols may be related with a fuzzy equivalence (called a
\emph{similarity}), these operations can be formally extended to
tolerate mismatches on functor names and/or arity or argument order. We
reformulate and extend previous work with a declarative approach
defining unification and generalization as sets of axioms and rules
forming a complete constraint-normalization proof system. These include
the Reynolds-Plotkin term-gene\-ra\-li\-za\-tion procedures, Maria
Sessa's ``weak'' unification with partially fuzzy signatures and its
corresponding generalization, as well as novel extensions of such
operations to fully fuzzy signatures (\ie, similar functors with
possibly different arities). One advantage of this approach is that it
requires no modification of the conventional data structures for terms
and substitutions. This and the fact that these declarative
specifications are efficiently executable conditional Horn-clauses
offers great practical potential for fuzzy infor\-mation-hand\-ling
applications.\footnote{This article appears in the pre-proceedings of
  LOPSTR 2017 with the title ``\textit{Lattice Operations on Terms with
    Fuzzy Signatures}.'' Its new title is technically more accurate.
The work presented in this paper is part of a
  wider
  study~\Cite{hak:gp:2017}{http://hassan-ait-kaci.net/pdf/fuzlatdks.pdf}.
  All proofs and more examples can be found in a more detailed
  paper~\Cite{hakpasi:fuzfotlat:2017}{http://hassan-ait-kaci.net/pdf/fuzfotlat.pdf}.}
\end{abstract}


\Section{Subsumption Lattice}
\label{fotlatticesection}

The first-order term (\fot) was introduced as a data structure in
software programming by the
\href{https://en.wikipedia.org/wiki/Prolog}{Prolog}
language.\flink{https://en.wikipedia.org/wiki/Prolog} Just like the
S-expression for LISP, the {\fot} is Prolog's universal data structure.
Using formal algebra notation, we write $\T_{\SubS{\Sigma},\V}$ for the
set of {\fots} on an operator signature $\Sigma \eqd
\bigcup_{n\geq0}\Sigma_n$ where $\Sigma_n$ is a set of operator symbols
of $n$ arguments $\Sigma_n \eqd \{f\,|\,\arity(f)=n, \, n\in\NN\}$, and
$\V$ is a set of variables.\footnote{We shall use Prolog's convention of
  writing variables with capitalized symbols.}  We shall designate an
element $f$ in $\Sigma$ as a \emph{functor}, with $\arity(f)$ denoting
its number of arguments.\footnote{When $\arity(f)=n$, this is often
  denoted by writing $f/n$.}  This set $\T_{\SubS{\Sigma},\V}$ can then
be defined inductively as:
\[
\T_{\SubS{\Sigma},\V} \; \eqd \; \V \; \cup \;
\{\mbox{$f(t_1,\ldots,t_n)$} \;|\; \mbox{$f\in\Sigma_n$},\;
\mbox{$t_i$}\in\T_{\SubS{\Sigma},\V},\; \mbox{$0\leq i\leq n, \;
  n\geq0$}\}.
\]
We write $c$ instead of $c()$ for a constant $c\in\Sigma_0$. Also, when
the set $\Sigma$ of functor symbols and the set $\V$ of variables are
implicit from the context, we simply write $\T$ instead of
$\T_{\SubS{\Sigma},\V}$. The set $\var(t)$ of variables
occurring in a {\fot} $t\in\T$ is defined as:
\[
\var(t) \; \eqd \; \left\{\begin{array}{l@{\;\;\mbox{if}\;\;}l}
                          \{ X \} & t=X\in\V 
                          \\[1.6ex]
                          \bigcup_{i=1}^n\var(t_n) & t=f(t_1,\ldots,t_n).
                          \end{array}
                   \right.
\]

\begin{figure}
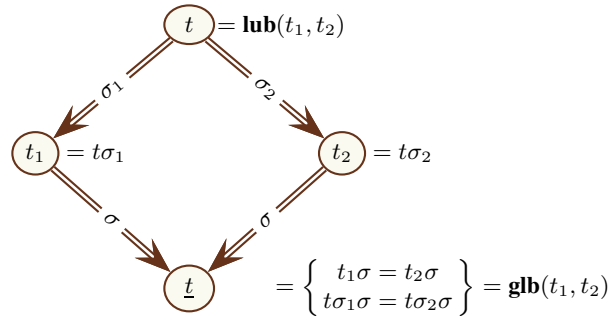

\newcommand{\NODE}[2]
{\ovalnode[fillstyle=solid,fillcolor=nodecolor,linecolor=brown]{#1}{#2}}
\newcommand{\SUBT}{\left\{
  \begin{array}{c}
    t_1\sigma = t_2\sigma \\
    t\sigma_1\sigma = t\sigma_2\sigma
  \end{array}
  \right\}
  }
\vspace{-6ex}
\begin{center}
\footnotesize
\begin{tabular}{c@{\hspace{.5cm}}c@{\hspace{1.3cm}}cl}
    \\
                                  & $\NODE{top}{\;t\;}$             &\hspace{-3cm}$= \lub(t_1,t_2)$
    \\ [1.2cm]
    $\NODE{t1}{t_1} = t\sigma_1$  &                                 &\hspace{-.5cm}$\NODE{t2}{t_2} = t\sigma_2$ 
    \\ [1.2cm]
                                  & $\NODE{bot}{\;\underline{t}\;}$ &\hspace{-.5cm}$= \SUBT$  & $= \glb(t_1,t_2)$
\end{tabular}

\psset{linecolor=brown,arrowsize=2pt 3,nrot=:U}

\ncline[doubleline=true]{<-}{t1}{top}\ncput*[nrot=:D]{\rotatebox{180}{$\sigma_1$}}
\ncline[doubleline=true]{<-}{t2}{top}\ncput*[nrot=:D]{$\sigma_2$}

\ncline[doubleline=true]{<-}{bot}{t1}\ncput*[nrot=:D]{$\sigma$}
\ncline[doubleline=true]{<-}{bot}{t2}\ncput*[nrot=:D]{\rotatebox{180}{$\sigma$}}

\end{center}

    \vspace{-6ex}
\caption{\textbf{Subsumption lattice operations}}
\label{fottermsub}
\end{figure}

The lattice-theoretic properties of {\fots} as data structures were
first exposed and studied by Reynolds
(in~\Cite{jcr:mi5:70}{http://www.cs.cmu.edu/afs/cs/user/jcr/ftp/transysalg.pdf})
and Plotkin (in~\cite{gdp:subsumption:70}
and~\Cite{gdp:mi5:70}{http://homepages.inf.ed.ac.uk/gdp/publications/MI5_note_ind_gen.pdf}).
They noted that the set $\T$ is ordered by term subsumption (denoted as
`$\issubsumedby$'); \viz, $t\issubsumedby t'$ (and we say: ``$t'$
\emph{subsumes} $t$'') iff there exists a variable substitution
$\sigma:\var(t')\arrow\T$ such that $t'\sigma = t$.  Two {\fots} $t$ and
$t'$ are considered ``\emph{equal up to variable renaming}'' (denoted as
$t\simeq t'$) whenever both $t\issubsumedby t'$ and $t'\issubsumedby
t$. Then, the set of first-order terms modulo variable renaming, when
lifted with a bottom element $\bot$ standing for ``\emph{no term}''
(\ie, the set $\T_{/\simeq}\cup\{\bot\}$) has a lattice structure for
subsumption. It has a top element $\top=\V$ (indeed, since any variable
in $\V$ can be substituted for any term, $\V$ is therefore the class of
any variable modulo renaming).  Unification corresponds to its greatest
lower bound (\glb) operation.  The dual operation, generalization of two
terms, yields a term that is their least upper bound (\lub) for
subsumption.  This can be summarized as the lattice diagram shown in
Fig.~\ref{fottermsub}.  In this diagram, given a pair of terms
$\couple{\leftfot}{\rightfot}$, the pair of substitutions
$\couple{\sigma_1}{\sigma_2}$ are their respective most general
generalizers, and the substitution $\sigma$ is the pair's most general
unifier (\mgu).  We formalize next these lattice operations on {\fots}
as declarative constraint normalization rules.
\vspace{-1cm}

\Subsection{Unification rules}
\label{fotunif}

In Fig.~\ref{hmmunifrules}, we give the set of equation normalization
rules that we shall call Herbrand-Martelli-Montanari
(\cite{herbrand:phd:1930}
and~\Cite{MandM}{http://moscova.inria.fr/~levy/courses/X/IF/03/pi/levy2/martelli-montanari.pdf}).
Each rule is \emph{provably correct} in that it is a solution-preserving
transformation of a set of equations.
\begin{figure}
\begin{center}\footnotesize
  \begin{Rules}
    \Urule[n\geq0]%
        {Term Decomposition}
        {E \cup \{\uni{f(s_1,\ldots,s_n)}{f(t_1,\ldots,t_n)}\}}
        {E \cup \{\uni{s_1}{t_1},\ldots,\uni{s_n}{t_n}\}}
    &
    \Urule%
        {Variable Erasure}
        {E \cup \{ \uni{X}{X} \}}
        {E}
    \\ \\
    \Urule[X\;\mbox{occurs in}\;E]%
        {Variable Elimination}
        {E \cup \{\uni{X}{t}\}}
        {E[X{\!\leftarrow\!}t] \cup \{\uni{X}{t}\}}
    &
    \Urule[t\not\in\V]%
        {Equation Orientation}
        {E \cup \{ \uni{t}{X} \}}
        {E \cup \{ \uni{X}{t} \}}
  \end{Rules}
\end{center}
\vspace{-2.5ex}
\caption{\textbf{Herbrand-Martelli-Montanari unification rules}}
\label{hmmunifrules}
\end{figure}
We can use these rules to unify two {\fots} $t_1$ and $t_2$.  We start
with the singleton set of equations $E \eqd \{\uni{t_1}{t_2}\}$, and
apply any applicable rule in any order until none applies. This always
terminates into a finite set of equations $E'$. If all the equations in
$E'$ are of the form $\uni{X}{t}$ with $X$ occurring nowhere else in
$E'$, then this is a most general unifying substitution (up to
consistent variable renaming) $\sigma \eqd \{~t/X \;|\; \uni{X}{t}\in
E'~\}$ solving the original equation (\ie, $t_1\sigma=t_2\sigma$);
otherwise, there is no solution---\ie, $\glb(t_1,t_2)=\bot$.
In these rules, we do not bother checking for circular terms
(``\emph{occurs-check}''). It can be done if wished; without it,
technically, these rules perform rational term
unification~\cite{jaffar:ngc:1984}.

\Subsection{Generalization rules}
\label{fotgen}

In 1970, John Reynolds and Gordon Plotkin published each an article, in
the same volume
(\Cite{jcr:mi5:70}{http://www.cs.cmu.edu/afs/cs/user/jcr/ftp/transysalg.pdf}
and~\Cite{gdp:mi5:70}{http://homepages.inf.ed.ac.uk/gdp/publications/MI5_note_ind_gen.pdf}),
giving two identical algorithms (up to notation) for the generalization
of two {\fots}.  Each describes a procedural method computing the most
specific {\fot} subsuming two given {\fots} in finitely many steps by
comparing them simultaneously, and generating a pair of generalizing
substitutions from a fresh variable wherever they disagree being scanned
from left to right, each time replacing the disagreeing terms by the new
variable everywhere they both occur in each term.


Next, we present a set of declarative normalization rules for
generalization which are equivalent to these procedural algorithms. As
far as we know, this is the first such presentation of a declarative set
of rules for generalization besides its more general form as
order-sorted feature term generalization
in~\Cite{ecml:2001}{http://www.hassan-ait-kaci.net/pdf/ecml01.pdf}.  The
advantage of specifying this operation in this manner rather than
procedurally as done originally by Reynolds and Plotkin is that each
rule or axiom relates a pair of prior substitutions to a pair of
posterior substitutions based only on local syntactic-pattern properties
of the terms to generalize, and this without resorting to side-effects
on global structures.  In this way, the terms and substitutions involved
are derived as solutions of logical syntactic constraints. In addition,
correctness of the so-specified operation is made much easier to
establish since we only need to prove each rule's correctness
independently of that of the others. Finally, the rules also provide an
effective means for the derivation of an operational semantics for the
so-specified operation by constraint solving, without need for control
specification as any applicable rule may be invoked in any
order.\footnote{Such as the Herbrand-Martelli-Montanari unification
  rules {\wrt} to Robinson's procedural unification algorithm.}

\begin{definition}[Generalization Judgement]\label{fotgenjudgement}
A \emph{generalization ju\-d\-gement} is an expression of the form:
\begin{equation}\label{fotgenjudgementexp}
\FotGenJudgement{\sigma_1}{\sigma_2}{t_1}{t_2}{t}{\theta_1}{\theta_2}
\end{equation}
where $\sigma_i:\var(t_i)\rightarrow\T$ and
$\theta_i:\var(t)\rightarrow\T$ ($i = 1,2$) are substitutions, and
$t\in\T$ and $t_i\in\T$ ($i = 1,2$) are {\fots}.
\end{definition}

\begin{definition}[Generalization Judgement Validity]\label{fotgenjudgementvalidity}
A generalization ju\-d\-gement such as~\emph{(\ref{fotgenjudgementexp})}
is said to be \emph{valid} whenever $t_i\sigma_i\,=\,t\theta_i,
\;\;\mbox{for}\; i = 1,2$.
\end{definition}


Contrary to other normalization rules in this document which are
expressed as conditional rewrite rules whereby a prior form (the
``numerator'') is related to a posterior form (the ``denominator''),
these normalization rules are more naturally rendered as (conditional)
Horn clauses of ju\-d\-gements.  This is as convenient as rewrite rules
since a Prolog-like operational semantics can then readily provide an
effective interpretation. This operational semantics is efficient
because it does not need backtracking as long as the complete set of
conditions of a ruleset covers all but mutually exclusive syntactic
patterns. Thus, a generalization rule is of the form:
\begin{equation}\label{fotgenrule}
\begin{array}{l}
  {[\phi]}
  \\
  \begin{array}{c}
    J_1 \hspace{.5cm}\ldots\hspace{.5cm} J_n
    \\ \hline
    J
  \end{array}
\end{array}
\end{equation}
where $\phi$ is a side meta-condition, and $J, J_1, \ldots, J_n$ are
ju\-d\-gements, and it reads, ``\emph{whenever the side condition $\phi$
  holds, if all the $n$ antecedent ju\-d\-gements $J_n$ are valid, then
  the consequent ju\-d\-gement $J$ is also valid}.''  Such a
generalization rule without a specified antecedent (a ``numerator'') is
called a ``\emph{generalization axiom}.''  Such an axiom is said to be
valid iff its consequent (the ``denominator'') is valid whenever its
optional side condition holds. It is equivalent to a rule where the
only antecedent is the trivial generalization ju\-d\-gement $\TRUE$.

\begin{definition}[Generalization Rule Correctness]\label{fotgenrule-correctness}
A conditional Horn rule such as Rule~(\ref{fotgenrule}) is
\emph{correct} iff $J_k$ is a valid judgement for all $k=1,\ldots,n$
implies that $J$ is a valid judgement, whenever the side condition
$\phi$ holds.
\end{definition}

Given $t_1$ and $t_2$ two {\fots} having no variable in common, in order
to find the most specific term $t$ and most general substitutions
$\sigma_i$, $i = 1,2$, such that $t\sigma_i=t_i$, $i = 1,2$, one needs
to establish the generalization ju\-d\-gement:
\begin{equation}\label{gengoal}
\FotGenJudgement{\empty}{\empty}{t_1}{t_2}{t}{\sigma_1}{\sigma_2}.
\end{equation}
In other words, this expresses the upper half of Fig.~\ref{fottermsub}
whereby $t\,=\,\lub(t_1,t_2)$, with most general substitutions
$\sigma_1$ and $\sigma_2$.  We give a complete set of normalization
axioms and rule for generalization for all syntactic patterns in
Fig.~\ref{fotgenrules}.
\begin{figure}
\footnotesize
\begin{Rules}
\GenAxiom%
        {Equal Variables}
        {\FotGenJudgement{\sigma_1}{\sigma_2}{X}{X}{X}{\sigma_1}{\sigma_2}}
& \hspace{-7cm}
\GenAxiom[t_1\in\V \;\mbox{or}\; t_2\in\V; \;\; t_1\neq t_2;\;\; X \;\mbox{is new}]
        {Variable-Term}
        {\FotGenJudgement{\sigma_1}{\sigma_2}{t_1}{t_2}{X}{\{\subst{t_1}{X}\}\sigma_1}{\{\subst{t_2}{X}\}\sigma_2}}
\\
\GenAxiom[m\geq0, n\geq0; \;\;m\neq n\;\mbox{or}\;f\neq g;\;\; X \;\mbox{is new}]
        {Unequal Functors}
        {\FotGenJudgement{\sigma_1}{\sigma_2}
                         {f(s_1,\ldots,s_m)}{g(t_1,\ldots,t_n)}
                         {X}
                         {\{\subst{f(s_1,\ldots,s_m)}{X}\}\sigma_1}
                         {\{\subst{g(t_1,\ldots,t_n)}{X}\}\sigma_2}}
\\
\GenRule[n\geq0]
       {Equal Functors}
       {\FotGenUnapplyJudgement{\sigma_1}{\sigma_2}{s_1}{t_1}{u_1}{\sigma^1_1}{\sigma^1_2}
        \hspace{.25cm}\ldots\hspace{.25cm}
        \FotGenUnapplyJudgement{\sigma^{n-1}_1}{\sigma^{n-1}_2}{s_n}{t_n}{u_n}{\sigma^n_1}{\sigma^n_2}}
       {\FotGenJudgement{\sigma_1}{\sigma_2}
                        {f(s_1,\ldots,s_n)}{f(t_1,\ldots,t_n)}
                        {f(u_1,\ldots,u_n)}
                        {\sigma^n_1}{\sigma^n_2}}
\end{Rules}
\vspace{-3ex}
\caption{\textbf{Generalization axioms and rule}}
\label{fotgenrules}
\end{figure}
Rule ``\RuleName{Equal Functors}'' uses an ``\emph{unapply}'' operation
(`$\,\unapply\,$') on a pair of terms $(t_1,t_2)$ given a pair of
substitutions $(\sigma_1,\sigma_2)$. It may be conceived as (and in fact
is) the result of simultaneously ``\emph{unapplying}'' $\sigma_i$ from
$t_i$ into a common variable $X$ only if such $X$ is bound to $t_i$ by
$\sigma_i$, for $i = 1,2$. If there is no such a variable, it is the
identity. Formally, this is defined as:
\begin{equation}\label{unapply}
  \stack{t_1}{t_2}\unapply\stack{\sigma_1}{\sigma_2} \;\eqd\;
  \left\{\begin{array}{ll}
  \stack{X}{X} & \mbox{if}\; t_i=X\sigma_i, \;\mbox{for}\; i = 1,2;
  \\ \\
  \stack{t_1}{t_2} & \mbox{otherwise}.
  \end{array}\right.
\end{equation}

Note also that Rule ``\RuleName{Equal Functors}'' is defined for
$n\geq0$. For $n=0$ (for any constant $c$), it becomes the following
axiom:
\begin{equation}\label{equal-constants-axiom}
\FotGenJudgement{\sigma_1}{\sigma_2}{c}{c}{c}{\sigma_1}{\sigma_2}.
\end{equation}

\begin{theorem}\label{theorem1}
The axioms and the rule of Fig.~\ref{fotgenrules} are correct.
\end{theorem}

In particular, with empty prior substitutions, we obtain the following corollary.
\begin{corollary}[{\fot} Generalization]\label{fotgencorollary}
Whenever the ju\-d\-gement
$\FotGenJudgement{\empty}{\empty}{t_1}{t_2}{t}{\sigma_1}{\sigma_2}$ is
valid, then $t\sigma_i=t_i$, for $i = 1,2$.
\end{corollary}

\Section{Fuzzy Lattice Operations}
\label{fuzzy-fot-lattice}

\Subsection{Fuzzy unification}
\label{fuzzy-fot-unification}


A fuzzy unification operation on {\fots}, dubbed ``\emph{weak
  unification},'' was proposed by Maria Sessa
in~\Cite{sessa:tcs:2002}{http://www.sciencedirect.com/science/article/pii/S0304397501001888}.
It normalizes equations between conventional {\fots} modulo a similarity
relation $\sim$ over functor symbols.  This similarity relation is then
homomorphically extended to one over all {\fots}. It is: (1)~the (crisp)
identity relation on variables (\ie, $X\sim_1X$, for any $X$ in $\V$);
otherwise, (2)~zero when either of the two terms is a variable (\ie,
$X\sim_0 t$ and $t\sim_0 X$, for any $X\neq t$ in $\V$); otherwise~(3):
\[
f(s_1,\ldots,s_n) \, \sim_{(\alpha\wedge\bigwedge_{i=1}^{n}\alpha_i)} \,
g(t_1,\ldots,t_n) \;\; \mbox{if} \;\; f\sim_{\alpha}g \; \mbox{and} \;
s_i\sim_{\alpha_i}t_i, \;\; i=1,\ldots,n
\]
where $\alpha\in[0,1]$ and $\alpha_i\in[0,1]$ $(i=1,\ldots,n)$ denote
the \textsl{unification degrees} to which each corresponding equation
holds.\footnote{The $\wedge$ operation used by Sessa in this expression
is $\min$; but other interpretations are
possible~(\Cite{fss:dubois:prade:1980}{ftp://ftp.micronet-rostov.ru/linux-support/books/computer\%20science/Fuzzy\%20systems/Fuzzy\%20Sets\%20And\%20Systems\%20Theory\%20And\%20Applications\%20-\%20Didier\%20Dubois\%20,\%20Henri\%20Prade.pdf}, \Cite{hak:gp:2017}{http://hassan-ait-kaci.net/pdf/fuzlatopdks.pdf}).}

In Fig.~\ref{mariasessaunification}, we provide a set of declarative
rewrite rules equivalent to Sessa's case-based ``weak unification
algorithm''~\Cite{sessa:tcs:2002}{http://www.sciencedirect.com/science/article/pii/S0304397501001888}.
To simplify the presentation of these rules while remaining faithful to
Sessa's weak unification algorithm, it is assumed for now that functor
symbols $f/m$ and $g/n$ of different arities $m\neq n$ are never
similar.
 This is without any loss of generality since Sessa's weak unification
 fails on term structures of different arities.\footnote{See Case~(2) of
   the weak unification algorithm given
   in~\Cite{sessa:tcs:2002}{http://www.sciencedirect.com/science/article/pii/S0304397501001888},
   Page~413.} Later, we will relax this and allow functors of different
 arities to be similar.  Note also that we do not bother checking for
 circular terms---but this can be done if wished.

\begin{figure}
\vspace{-.5cm}
\begin{center}\footnotesize
  \begin{Rules}
    \Urule[\stackanchor{$f\sim_{\beta}g$}{$n\geq0$}]%
        {Fuzzy Term Decomposition}
        {(E \cup \{ \uni{f(s_1,\ldots,s_n)}{g(t_1,\ldots,t_n)} \})_\alpha}
        {(E \cup \{ \uni{s_1}{t_1}, \ldots, \uni{s_n}{t_n} \})_{\alpha\wedge\beta}}
  &
        \hspace{-.65cm}
        \Urule%
        {Variable Erasure}
        {(E \cup \{ \uni{X}{X} \})_\alpha}
        {E_\alpha}
  \\
   \Urule[X\;\mbox{occurs in}\;E]%
        {Variable Elimination}
        {(E \cup \{ \uni{X}{t} \})_\alpha}
        {(E[X\!\leftarrow\!t] \cup \{ \uni{X}{t} \})_\alpha}
  &
        \hspace{-.65cm}
        \Urule[t\not\in\V]%
        {Equation Orientation}
        {(E \cup \{ \uni{t}{X} \})_\alpha}
        {(E \cup \{ \uni{X}{t} \})_\alpha}
  \end{Rules}
\end{center}
\vspace{-.5cm}
\caption{\textbf{Normalization rules corresponding to Maria Sessa's
    ``weak unification''}}
\label{mariasessaunification}
\end{figure}

The rules of Fig.~\ref{mariasessaunification} transform $E_\alpha$ a
finite conjunctive set $E$ of equations among {\fots} along with an
associated truth value, or ``\emph{unification degree},''
$\alpha\in[0,1]$, into $E'_{\alpha'}$ another set of equations $E'$ with
truth value $\alpha'\in[0,\alpha]$. Given to solve a fuzzy
unification equation $\uni{s}{t}$ between two {\fots} $s$ and $t$, form
the set $\{
\uni{s}{t} \}_1$ (\ie, with unification degree 1), and apply any
applicable rules in Fig.~\ref{mariasessaunification} until either the
unification degree of the set of equations is $0$ (in which case there
is no solution to the original equation, not even a fuzzy one), or the
final resulting set $E_\alpha$ is a solution with truth value $\alpha$
in the form of a variable substitution $\sigma\eqd \{\subst{X}{t} \;|\;
\uni{X}{t}\in E\}$ such that $s\sigma \sim_\alpha t\sigma$.


From our perspective, a fuzzy unification operation ought to be able to
fuzzify \emph{full} {\fot} unification: whether (1)~functor symbol
mistmatch, and/or (2)~arity mismatch, and/or (3)~in which order subterms
correspond.  Sessa's fuzzification of unification as weak unification
misses on the last two items.  This is unfortunate as this can turn out
to be quite useful. In real life, there is indeed no such garantee that
argument positions of different functors match similar information in
data and knowledge bases, hence the need for
alignment~\Cite{SiGMa:kbalign:2013}{http://snap.stanford.edu/social2012/papers/lacostejulien-palla-etal.pdf}.

Still, it has several qualities:
\begin{itemize}
  \item \emph{It is simple}---specified as a straightforward extension
    of crisp unification: only one rule (Rule ``\RuleName{Fuzzy Term
      Decomposition}'') may alter the fuzziness of an equation set by
    tolerating similar functors.

  \item \emph{It is conservative}---neither {\fots} nor {\fot}
    substitutions \emph{per se} need be fuzzified; so conventional crisp
    representations and operations can be used; if restricted to only
    $0$ or $1$ truth values, it is equivalent to crisp {\fot}
    unification.
\end{itemize}

We now give an extension of Sessa's weak unification which can tolerate
such fuzzy similarity among functors of different arities.  Given a
similarity relation $\sim$ on a ranked signature $\Sigma \,\eqd\,
\Sigma_{n\geq 0}$, $\sim:\Sigma^2\arrow[0,1]$ which, unlike M.~Sessa's
equal-arity condition, now allows mismatches of similar symbols with
distinct arities or equal arities but different argument orders. Namely,
\begin{itemize}

\item it admits that $(\sim\,\cap\;\Sigma_m\times\Sigma_n) \;\neq\;
  \empty$ for some $m\geq0$, $n\geq0$, such that $m\neq n$;

\item for each pair of functors $\couple{f}{g}\,\in\Sigma^2$, such that
  $f\,\in\Sigma_m$ and $g\,\in\Sigma_n$, with $0\leq m\leq n$, and
  $f\sim_\alpha g$, $(\alpha\in(0,1])$, there exists an injective (\ie,
  one-to-one) mapping $p:\{1,\ldots,m\}\arrow\{1,\ldots,n\}$ associating
  each of the $m$ argument positions of $f$ to a unique position among
  the $n$ arguments of $g$ (which is denoted as $f\sim_\alpha^p g$).
\end{itemize}
Note that in the above, $m$ and $n$ are such that $0\leq m\leq n$; so
the one-to-one argument-position mapping goes from the lesser set to the
larger set. There is no loss of generality with this assumption as this
will be taken into account in the normalization rules.


\begin{Example}[Similar functors with different arities]\label{unequalaritysimilarity}
Consider $\Person/3$, a functor of arity 3, and $\Individual/4$, a
functor of arity 4 with:
\begin{itemize}
\item similarity truth value of $.9$; \ie,
  $\texttt{person}/3\;\ \sim_{.9}\;\;\texttt{individual}/4$; and,
\item one-to-one position mapping $p:\{1,2,3\}\arrow\{1,2,3,4\}$:
  \[
  \mbox{from}\; \Person/3 \; \mbox{to}\; \Individual/4 \; \mbox{with} \;
  p \,:\, \{1\arrow1, 2\arrow3, 3\arrow4 \}
  \]
  so that:
  \[
  \texttt{person}(\textit{Name},\textit{SSN},\textit{Address})
  \;\;\sim^p_{.9}\;\;
  \Individual(\textit{Name},\textit{DoB},\textit{SSN},\textit{Address})
  \]
\end{itemize}
writing $f\sim^p_\alpha g$ a similarity relation between a functor $f$
and a functor $g$ of truth value $\alpha$ and $f$-to-$g$
argument-position mapping $p$; in our example, $\Person
\sim^{\{1\arrow1,2\arrow3,3\arrow4\}}_{.9} \Individual$.

With this kind of specification, we can tolerate not only fuzzy
mismatching of terms with distinct functors \texttt{person} and
\texttt{individual}, but also up to a correspondance of argument
positions from \texttt{person} to {\Individual} specified as $p$,
all with a truth value of $.9$.

\end{Example}


Starting with the Herbrand-Martelli-Montanari ruleset of
Fig.~\ref{hmmunifrules}, fuzziness is introduced by relaxing
``\RuleName{Term Decomposition}'' to make it also tolerate possible
arity or argument-order mistmatch in two structures being unified. In
other words, the given functor similarity relation $\sim$ is adjoined a
position mapping from argument positions of a functor $f$ to those of a
functor $g$ when $f\neq g$ and $f\sim_\alpha g$ with $\alpha\in(0,1]$.
  This is then taken into account in tolerating a fuzzy mismatch between
  two term structures $s=f(s_1,\ldots,s_m)$ and
  $t=g(t_1,\ldots,t_n)$. This may involve a mismatch between the terms'
  functor symbols ($f$ and $g$), their arities ($m$ and $n$), subterm
  orders, or a combination. We first reorient all such equations by
  flipping sides so that the left-hand side is the one wih lesser or
  equal arity.  In this manner, assuming $f\sim^p_\beta g$ and
  $0\leq\alpha,\beta\leq1$, an equation of the form:
$\bigl\{\uni{f(s_1,\ldots,s_m)}{g(t_1,\ldots,t_n)}\bigr\}_\alpha$
for $0\leq m\leq n$ acquires its truth value $\alpha\wedge\beta$ due to
functor and arity mismatch when equated.  A fully fuzzified
term-decomposition rule should proceed with replacing such a fuzzy
structure equation with the following conjunction of fuzzy equations
between subterms at corresponding indices given by the one-to-one argument mapping $p:\{1,\ldots,m\}\arrow\{1,\ldots,n\}$:
$\bigl\{\uni{s_1}{t_{p(1)}},
\;\;\ldots,\;\;
\uni{s_m}{t_{p(m)}}, \;\;\ldots\;\; \bigr\}_{\alpha\wedge\beta}$.
Note that all the subterms in the right-hand side term that are
arguments at indices which are not $p$-images are ignored as they have
no counterparts in the left-hand side.  These terms are simply dropped
as part of the fuzzy approximative unification.  This generic
rule is shown in Fig.~\ref{fuzzytermdecomposition} along with another
rule needed to make it fully effective: a rule reorienting a term
equation into one with a lesser-arity term on the left.
\begin{figure}
\vspace{-.25cm}
\begin{center}\small
  \renewcommand{\gstick}{\rule[-1.5ex]{0pt}{4ex}}
  \begin{Rules}
    \LeftGenRule[0\leq m\leq n;\; f\sim^p_\beta g]%
        {Generic Weak Term Decomposition}
        {\left(E \cup \{ \uni{f(s_1,\ldots,s_m)}{g(t_1,\ldots,t_n)} \}\right)_{\alpha}\hspace{-1.2cm}}
        {\left(E \cup \{ \uni{s_{1}}{t_{p(1)}}, \ldots, \uni{s_{m}}{t_{p(m)}}\}\right)_{\alpha\wedge\beta}\hspace{-1.2cm}}
\\ 
    \LeftGenRule[0\leq n < m]%
        {Fuzzy Equation Reorientation}
        {\left(E \cup \{ \uni{f(s_1,\ldots,s_m)}{g(t_1,\ldots,t_n)} \}\right)_{\alpha}}
        {\left(E \cup \{ \uni{g(t_1,\ldots,t_n)}{f(s_1,\ldots,s_m)} \}\right)_{\alpha}}
  \end{Rules}
  \renewcommand{\gstick}{\rule[-3ex]{0pt}{7ex}}
\end{center}
\vspace{-.5cm}
\caption{\textbf{Generic fuzzification of {\fot} unification's decomposition rule}}
\label{fuzzytermdecomposition}
\end{figure}

\vspace{-.5cm}
\begin{theorem}\label{theorem2}
The fuzzy unification rules of Fig.~\ref{mariasessaunification} where
Rule ``\RuleName{Fuzzy Term Decomposition}'' is replaced by the rules of
Fig.~\ref{fuzzytermdecomposition} are correct.
\end{theorem}

In other words, applying this modified ruleset to
$E_1\eqd\{\uni{s}{t}\}_1$, an equation set of truth value $1$ (in any
order as long as a rule applies and its truth value is not zero) always
terminates. And when the final equation set is a substitution $\sigma$,
it is a fuzzy solution with truth value $\alpha$ such that
$s\sigma\sim_\alpha t\sigma$.

\begin{Example}[{\fot} fuzzy unification with similar functors of different arities]\label{fuzvarargunifex}
  Let us take a functor signature such that:
  $\{a,b,c,d\}\subseteq\Sigma_0$,
  $\{f,g,\ell\}\subseteq\Sigma_2$,
  $\{h\}\subseteq\Sigma_3$;
  and let us further assume that the only non-zero similarities argument
  mappings among these functors are:
  \begin{itemize}
  \item $a\sim_{.7}b$,
  \item $c\sim_{.6}d$,
  \item $f\sim^{\{1\arrow2,2\arrow1\}}_{.9}g$ and $g\sim^{\{1\arrow2,2\arrow1\}}_{.9}f$,
  \item $\ell\sim^{\{1\arrow2,2\arrow3\}}_{.8}h$.
  \end{itemize}

Let us consider the fuzzy equation set $\{\uni{\leftfot}{\rightfot}\}_1$:
\begin{equation}\label{abcdfghex2}
\left\{
    {\red\uni{h(X,g(Y,b),f(Y,c))}{\ell(f(a,Z),g(d,c))}}
\right\}_1
\end{equation}
and let us apply the rules of Figure~\ref{mariasessaunification}
with rule \RuleName{Weak Term Decomposition} is replaced by the
rules of Figure~\ref{fuzzytermdecomposition}:
\begin{itemize}
\item apply Rule \RuleName{Fuzzy Equation Reorientation} with $\alpha=1$
  since $\arity(\ell)<\arity(h)$:
\[
\left\{
     \uni{\ell(f(a,Z),g(d,c))}{h(X,g(Y,b),f(Y,c))}
\right\}_1;
\]

\item apply Rule \RuleName{Generic Weak Term Decomposition} to:
  {\red\[\uni{\ell(f(a,Z),g(d,c))}{h(X,g(Y,b),f(Y,c))}\]} with
  $\alpha=1$ and $\beta=.8$ since
  $\ell\sim^{\{1\arrow2,2\arrow3\}}_{.8}h$, to obtain:
\[
\left\{
     \uni{f(a,Z)}{g(Y,b)},
     \uni{g(d,c)}{f(Y,c)}
\right\}_{.8};
\]

\item apply Rule \RuleName{Generic Weak Term Decomposition} to
  {\red$\uni{f(a,Z)}{g(Y,b)}$} with $\alpha=.8$ and $\beta=.9$ since
  $f\sim^{\{1\arrow2,2\arrow1\}}_{.9}g$, to obtain:
\[
\left\{
     \uni{a}{b},
     \uni{Z}{Y},
     \uni{g(d,c)}{f(Y,c)}
\right\}_{.8};
\]

\item apply Rule \RuleName{Generic Weak Term Decomposition} to
  {\red$\uni{a}{b}$} with $\alpha=.8$ and $\beta=.7$ since
  $a\sim_{.7}b$, to obtain:
\[
\left\{
     \uni{Z}{Y},
     \uni{g(d,c)}{f(Y,c)}
\right\}_{.7};
\]

\item apply Rule \RuleName{Generic Weak Term Decomposition} to
  {\red$\uni{g(d,c)}{f(Y,c)}$} with $\alpha=.7$ and $\beta=.9$ since
  $f\sim^{\{1\arrow2,2\arrow1\}}_{.9}g$, to obtain:
\[
\left\{
     \uni{Z}{Y},
     \uni{d}{c},
     \uni{c}{Y}
\right\}_{.7};
\]

\item apply Rule \RuleName{Generic Weak Term Decomposition} to
  {\red$\uni{d}{c}$} with $\alpha=.7$ and $\beta=.6$ since
  $d\sim_{.6}c$, to obtain:
\[
\left\{
     \uni{Z}{Y},
     \uni{c}{Y}
\right\}_{.6};
\]

\item apply Rule \RuleName{Equation Orientation} to
  {\red$\uni{c}{Y}$} with $\alpha=.6$, to obtain:
\[
\left\{
     \uni{Z}{Y},
     \uni{Y}{c}
\right\}_{.6}.
\]

\item apply Rule \RuleName{Variable Elimination} to
  {\red$\uni{Y}{c}$} with $\alpha=.6$, to obtain:
\[
\left\{
     \uni{Z}{c},
     \uni{Y}{c}
\right\}_{.6}.
\]
\end{itemize}
This last equation set is in normal form with truth value $.6$ and
defines the substitution $\sigma = \left\{\, \subst{c}{Z}, \subst{c}{Y}
\,\right\}$ so that:
\begin{equation}\label{fuzabcdfghex}
\leftfot\sigma = h(X,g(Y,b),f(Y,c))\left\{\,\subst{c}{Z}, \subst{c}{Y} \,\right\}
\;\sim_{.6}\; \rightfot\sigma = 
\ell(f(a,Z),g(d,c))\left\{\,\subst{c}{Z}, \subst{c}{Y} \,\right\},
\end{equation}
that is:
\begin{equation}\label{fuzabcdfghex1}
\leftfot\sigma = h(X,g(c,b),f(c,c))
\;\sim_{.6}\; \rightfot\sigma = 
\ell(f(a,c),g(d,c)).
\end{equation}
\end{Example}

\begin{Example}[The same fuzzy unification with more expressive symbols]\label{three-item-box-bag-ex}
Let us give more expressive names to functors of
Example~\ref{fuzvarargunifex} in the context of, say, a gift-shop Prolog
database which describes various configurations for multi-item gift
boxes or bags containing such items as flowers, sweets, \etc, which can
be already joined as pairs or not joined as loose couples.
\begin{itemize}
  \item $a \eqd \Violet$,
  \item $b \eqd \Lilac$,
  \item $c \eqd \Chocolate$,
  \item $d \eqd \Candy$,
  \item $f \eqd \Pair$,
  \item $g \eqd \Couple$,
  \item $\ell \eqd \SmallGiftBag$,
  \item $h \eqd \SmallGiftBox$,
\end{itemize}
with the following similarity degrees and argument mappings,:
\begin{itemize}
  \item $\Violet\sim_{.7}\Lilac$,
  \item $\Chocolate\sim_{.6}\Candy$,
  \item $\Pair\sim_{.9}\Couple$,
  \item $\Pair\sim^{\{1\arrow2,2\arrow1\}}_{.9}\Couple$ and $\Couple\sim^{\{1\arrow2,2\arrow1\}}_{.9}\Pair$,
  \item $\SmallGiftBag\sim^{\{1\arrow2,2\arrow3\}}_{.8}\SmallGiftBox$.
\end{itemize}

With these functors Equation~\eqref{abcdfghex2} now reads:
\begin{quote}
  \begin{TABULAR}{ll}
  $(\leftfot)$ & 
  \parbox{.25\textwidth}{\begin{tabbing}
       $\SmallGiftBox\;$\=$(\;$\=$\;X$ \\
                        \>$,$  \>$\;\Couple(Y,\Lilac)$ \\
                        \>$,$  \>$\;\Pair(Y,\Chocolate)$ \\
                        \>$)$
       \end{tabbing}}
  \\ [-1.5ex]
  & $\doteq$
  \\ [-1.5ex]
  $(\rightfot)$ & 
  \parbox{.25\textwidth}{\begin{tabbing}
       $\SmallGiftBag\;$\=$(\;$\=$\;\Pair(\Violet,Z)$ \\
                        \>$,$  \>$\;\Couple(\Candy,\Chocolate)$ \\
                        \>$)$
  \end{tabbing}}
  \end{TABULAR}
\end{quote}

With the new functor symbols, the substitution $\sigma = \left\{\,
\subst{\Chocolate}{Z}, \subst{\Chocolate}{Y} \,\right\}$ obtained after
normalization yields the fuzzy solution:
\begin{quote}
  \begin{TABULAR}{ll}
  $(\leftfot\sigma)$ &
  \parbox{.25\textwidth}{\begin{tabbing}
       $\SmallGiftBox\;$\=$(\;$\=$\;X$ \\
                        \>$,$  \>$\;\Couple(\Chocolate,\Lilac)$ \\
                        \>$,$  \>$\;\Pair(\Chocolate,\Chocolate)$ \\
                        \>$)$
       \end{tabbing}}
  \\ [-1.5ex]
  & $\sim_{.6}$
  \\ [-1.5ex]
  $(\rightfot\sigma)$ &
  \parbox{.25\textwidth}{\begin{tabbing}
       $\SmallGiftBag\;$\=$(\;$\=$\;\Pair(\Violet,\Chocolate)$ \\
                        \>$,$  \>$\;\Couple(\Candy,\Chocolate)$ \\
                        \>$)$
  \end{tabbing}}
   \end{TABULAR}
\end{quote}
with truth value $.6$ capturing the unification degree to which $\sigma$
solves the original equation.
\end{Example}



Rule \RuleName{Generic Weak Term Decomposition} is a very general
rule for normalizing fuzzy equations over {\fot} structures. It has the
following convenient properties:
\begin{enumerate}
\item it accounts for fuzzy mismatches of similar functors of possibly
  different arity or order of arguments;

\item when restricted to tolerating only similar equal-arity functors
  with matching argument positions, it reduces to Sessa's weak
  unification's \RuleName{Weak Term Decomposition} rule;

\item when truth values are further restricted to be in $\{0,1\}$, it
  reduces to Herbrand-Martelli-Montanari's \RuleName{Term Decomposition}
  rule;

\item it requires no alteration of the standard notions of {\fots} and
  {\fot} substitutions: similarity among {\fots} is derived from that of
  signature symbols;

\item finally, and most importantly, it keeps fuzzy unification in the same
  complexity class as crisp unification: that of Union-Find
  (~\Cite{knight:unification:1989}{http://citeseerx.ist.psu.edu/viewdoc/download;jsessionid=92AF7CA745E2C0B8EB619F09FFB5D3CA?doi=10.1.1.64.8967&rep=rep1&type=pdf}, \Cite{union:find:princeton:2013}{https://www.cs.princeton.edu/~wayne/kleinberg-tardos/pdf/UnionFind.pdf}).\footnote{Quasi-linear; \ie,
  linear with a $\log\ldots\log$ coefficient~\cite{ahu}.}
  
\end{enumerate}
As a result, it is more general than all other extant approaches we know
which propose a fuzzy {\fot} unification operation. The same will be
established for the fuzzification of the dual operation: first a limited
``\emph{functor-weak}'' {\fot} generalization corresponding to the dual
operation of Sessa's ``weak'' unification, then to a more expressive
``\emph{functor/arity-weak}'' {\fot} generalization corresponding to our
extension of Sessa's unification to functor/arity weak unification.


\Subsection{Fuzzy generalization}
\label{fuzzy-fot-generalization}


Let $t_1$ and $t_2$ be two {\fots} in $\T$ to generalize.  We shall use
the following notation for a fuzzy generalization ju\-d\-gement:
\begin{equation}\label{fuzfotgenjudgement}
\FFotGenJudgement{\sigma_1}{\sigma_2}{\alpha}{t_1}{t_2}{t}{\theta_1}{\theta_2}{\beta}
\end{equation}
given:
\begin{itemize}
\item $\sigma_i:\var(t_i)\rightarrow\T$ ($i = 1,2$): two prior
  substitutions with prior truth value $\alpha$,
\item $t_i$ ($i = 1,2$): two prior {\fots},
\item $t$: a posterior {\fot},
\item $\theta_i:\var(t)\rightarrow\T$ ($i = 1,2$): two posterior
  substitutions with truth value $\beta$.
\end{itemize}
\begin{definition}[Fuzzy Generalization Judgement Validity]\label{validfuzgenjudgement}
A fuzzy generalization ju\-d\-gement such
as~\emph{(\ref{fuzfotgenjudgement})} is \emph{valid} whenever
$0\leq\beta\leq\alpha\leq1$ and $t_i\sigma_i\sim_\beta t\theta_i$ for $i
= 1,2$.
\end{definition}

\begin{definition}[Fuzzy Generalization Rule Correctness]\label{fuzfotgenrulecorrectness}
A fuzzy generalization rule is correct iff, whenever the side condition
holds, if all the fuzzy generalization judgements making up its
antecedent are valid, then necessarily the generalization judgement in
its consequent is valid.
\end{definition}

In Fig.~\ref{fuzfotgenrules}, we give a fuzzy version of the
generalization rules of Fig.~\ref{fotgenrules}. As was the case in
Sessa's weak unification, we assume as well (for now) that we are only
given a similarity relation $\sim\;\in\,\Sigma\times\Sigma\arrow[0,1]$
on the signature $\Sigma=\cup_{n\geq 0}\Sigma_n$ such that for all
$m\geq0$ and $n\geq0$, $m\neq n$ implies
$\sim\,\cap\,\Sigma_m\times\Sigma_n\;=\;\empty$ (\ie, if functors $f$
and $g$ have different arities, then $f\not\sim g$).

\begin{figure}
\footnotesize
\vspace{-.25cm}
\begin{Rules}
\hspace{-.7cm}%
\GenAxiom%
       {Fuzzy Equal Variables}
       {\FFotGenJudgement{\sigma_1}{\sigma_2}{\alpha}{X}{X}{X}{\sigma_1}{\sigma_2}{\alpha}}
& \hspace{-8cm}
\GenAxiom[t_1\in\V \;\mbox{or}\; t_2\in\V; \;\; t_1\neq t_2; \;\; X \;\mbox{is new}]
       {Fuzzy Variable-Term}
       {\FFotGenJudgement{\sigma_1}{\sigma_2}{\alpha}{t_1}{t_2}{X}{\{\subst{t_1}{X}\}\sigma_1}{\{\subst{t_2}{X}\}\sigma_2}{\alpha}}
\\
\hspace{-.7cm}%
\GenAxiom[f\not\sim g;\; m\geq0, n\geq0;\; X \;\mbox{is new}]
       {Dissimilar Functors}
       {\FFotGenJudgement{\sigma_1}{\sigma_2}
                         {\alpha}
                         {f(s_1,\ldots,s_m)}{g(t_1,\ldots,t_n)}
                         {X}
                         {\{\subst{f(s_1,\ldots,s_m)}{X}\}\sigma_1}
                         {\{\subst{g(t_1,\ldots,t_n)}{X}\}\sigma_2}
                         {\alpha}}
\\
\hspace{-.7cm}%
\GenRule[f\sim_\beta g;\;\; n\geq0;\;\; \alpha_0\eqd\alpha\wedge\beta]
      {Similar Functors}
      {\FFotGenUnapplyJudgement{\sigma_1}{\sigma_2}{\alpha_0}{s_1}{t_1}{u_1}{\sigma^1_1}{\sigma^1_2}{\alpha_1}
       \,\ldots\,
       \FFotGenUnapplyJudgement{\sigma^{n-1}_1}{\sigma^{n-1}_2}{\alpha_{n-1}\!\!}{s_n}{t_n}{u_n}{\sigma^n_1}{\sigma^n_2}{\alpha_n}}
      {\FFotGenJudgement{\sigma_1}{\sigma_2}
                        {\alpha}
                        {f(s_1,\ldots,s_n)}{g(t_1,\ldots,t_n)}
                        {f(u_1,\ldots,u_n)}
                        {\sigma^n_1}{\sigma^n_2}
                        {\alpha_n}}
\end{Rules}
\caption{\textbf{Functor-weak generalization axioms and rule}}
\label{fuzfotgenrules}
\end{figure}

Rule \RuleName{Similar Functors} uses a ``\emph{fuzzy unapply}'' operation
(`$\,\fuzunapply{\alpha}$') on a pair of terms $(t_1,t_2)$ given a pair of
substitutions $(\sigma_1,\sigma_2)$ and a truth value $\alpha$. It is the result
of ``\emph{unapplying}'' $\sigma_i$ from $t_i$ into a common variable, if any,
whenever it is bound by $\sigma_1$ to a term $t'_1$ and by $\sigma_2$ to a term
$t'_2$ which are respectively $\alpha$-similar to $t_i$ for $i = 1,2$.  It is
defined as:
\begin{equation}\label{fuzzyunapply}
  \stack{t_1}{t_2}\fuzunapply{\alpha}\stack{\sigma_1}{\sigma_2} \;\eqd\;
  \left\{\begin{array}{ll}
  \stack{X}{X} & \mbox{if}\; t_i\sim_{\alpha}X\sigma_i \;\mbox{for}\; $i = 1,2$;
  \\ \\
  \stack{t_1}{t_2} & \mbox{otherwise}.
  \end{array}\right.
\end{equation}

\begin{theorem}\label{theorem3}
The fuzzy generalization rules of Fig.~\ref{fuzfotgenrules} are
correct.
\end{theorem}

\begin{Example}[{\fot} fuzzy generalization]\label{fuzfotgenex}
\footnotesize
Let us apply the fuzzy generalization axioms and rules
of Figure~\ref{fuzfotgenrules} to:
\[
\begin{ARRAY}{l@{\;\eqd\;}l}
\leftfot  & h(f(a,X_1),g(X_1,b),f(Y_1,Y_1)),
\\[2ex]
\rightfot & h(X_2,X_2,g(c,d)).
\end{ARRAY}
\]

\begin{itemize}
\item
Let us find term $t$, substitutions
  $\sigma_i:\var(t)\arrow\var(\textbf{t}_i)$ ($i=1,2$), and truth value
  $\alpha\in[0,1]$ such that $t\sigma_1\sim_\alpha
  h(f(a,X_1),g(X_1,b),f(Y_1,Y_1))$ and $t\sigma_2\sim_\alpha
  h(X_2,X_2,g(c,d))$; that is, solve the following fuzzy generalization
  constraint problem:
  \[
  \FFotGenJudgement{\empty}
                   {\empty}
                   {1}
                   {h(f(a,X_1),g(X_1,b),f(Y_1,Y_1))}
                   {h(X_2,     X_2,     g(c,  d))}
                   {t}
                   {\sigma_1}{\sigma_2}
                   {\alpha}.
   \]

   \item
   By Rule \RuleName{Similar Functors}, we must have
  $t=h(u_1,u_2,u_3)$ since:
  \[
  \FFotGenJudgement{\empty}
                   {\empty}
                   {1}
                   {h(f(a,X_1),g(X_1,b),f(Y_1,Y_1))}
                   {h(X_2,     X_2,     g(c,  d))}
                   {h(u_1,u_2,u_3)}
                   {\sigma_1}
                   {\sigma_2}
                   {\alpha}
   \]
   where:
   \begin{itemize}
   \item
   $u_1$ is the fuzzy generalization of
       $\stack{f(a,X_1)}{X_2}\fuzunapply{1}\stack{\empty}{\empty}$;
       that is, of $f(a,X_1)$ and $X_2$; and by Rule \RuleName{Fuzzy
         Variable-Term}: 
      \[
      \FFotGenJudgement{\empty}
                       {\empty}
                       {1}
                       {f(a,X_1)}
                       {X_2}
                       {X}
                       {\{ \subst{f(a,X_1)}{X} \}}
                       {\{ \subst{X_2}{X} \}}
                       {1}
       \]
       and so $u_1=X$;

       $u_2$ is the fuzzy generalization of
       $\stack{g(X_1,b)}{X_2}\fuzunapply{1}\stack{\{
       \subst{f(a,X_1)}{X} \}}{\{ \subst{X_2}{X}\} }$; that is, of $g(X_1,b)$ and
       $X_2$; and by Rule \RuleName{Fuzzy Variable-Term}:
      \[
      \FFotGenJudgement{\{ \subst{f(a,X_1)}{X} \}}
                       {\{ \subst{X_2}{X}\} }
                       {1}
                       {g(X_1,b)}
                       {X_2}
                       {Y}
                       {\{ \ldots, \subst{g(X_1,b)}{Y} \}}
                       {\{ \ldots, \subst{X_2}{Y} \}}
                       {1}
       \]
       and so $u_2=Y$;

       \item
       $u_3 = f(v_1,v_2)$ is the fuzzy generalization of
       \[\stack{f(Y_1,Y_1)}{g(c,d)}\fuzunapply{.9}\stack{\{
       \subst{f(a,X_1)}{X}, \subst{g(X_1,b)}{Y} \}} {\{ \subst{X_2}{X},
       \subst{X_2}{Y} \}};\] that is, of $f(Y_1,Y_1)$ and $g(c,d)$ with
       truth value $.9$, because of Rule \RuleName{Similar Functors} and
       $f\sim_{.9}g$, and:
       \begin{itemize}
       \item
       $v_1$ is the fuzzy generalization of \[\stack{\{
         \subst{f(a,X_1)}{X}, \subst{g(X_1,b)}{Y} \}}{\{ \subst{X_2}{X},
         \subst{X_2}{Y} \}}\fuzunapply{.9}\stack{Y_1}{c};\] that is, of
         $Y_1$ and $c$; and by Rule \RuleName{Fuzzy Variable-Term}:
         \[
         \FFotGenJudgement{\{ \subst{f(a,X_1)}{X}, \subst{g(X_1,b)}{Y} \}}
                          {\{ \subst{X_2}{X}, \subst{X_2}{Y} \}}
                          {.9}
                          {Y_1}
                          {c}
                          {Z}
                          {\{ \ldots, \subst{Y_1}{Z} \}}
                          {\{ \ldots, \subst{c}{Z}  \}}
                          {.9}
         \]
         that is, $v_1=Z$;
       \item
         $v_2$ is the fuzzy generalization of
         \[\stack{Y_1}{d}\fuzunapply{.9}\stack{\{ \subst{f(a,X_1)}{X},
         \subst{g(X_1,b)}{Y}, \subst{Y_1}{Z} \}}{\{ \subst{X_2}{X},
         \subst{X_2}{Y}, \subst{c}{Z} \}};\] that is, of $Y_1$ and $d$;
         and by Rule \RuleName{Fuzzy Variable-Term}:
         \[
         \FFotGenJudgement{\{ \subst{f(a,X_1)}{X}, \subst{g(X_1,b)}{Y}, \subst{Y_1}{Z} \}}
                          {\{ \subst{X_2}{X}, \subst{X_2}{Y}, \subst{c}{Z}  \}}
                          {.9}
                          {Y_1}
                          {d}
                          {U}
                          {\{ \ldots, \subst{Y_1}{U} \}}
                          {\{ \ldots, \subst{d}{U}   \}}
                          {.9}
         \]
         that is, $v_2=U$;
                       
       \end{itemize}
     in other words, $u_3=f(Z,U)$ since:
      \[
      \FFotGenJudgement{\{ \subst{f(a,X_1)}{X}, \subst{g(X_1,b)}{Y} \}}
                       {\{ \subst{X_2}{X}, \subst{X_2}{Y} \}}
                       {1}
                       {f(Y_1,Y_1)}
                       {g(c,d)}
                       {f(Z,U)}
                       {\{ \ldots, \subst{Y_1}{Z}, \subst{Y_1}{U} \}}
                       {\{ \ldots, \subst{c}{Z}, \subst{d}{U}   \}}
                       {.9}
       \]

   \end{itemize}
\end{itemize}
and so:
  \[
  \FFotGenJudgement{\empty}
                   {\empty}
                   {1}
                   {\leftfot}
                   {\rightfot}
                   {h(X,Y,f(Z,U))}
                   {\{ \subst{f(a,X_1)}{X}, \subst{g(X_1,b)}{Y}, \subst{Y_1}{Z}, \subst{Y_1}{U} \}}
                   {\{ \subst{X_2}{X}, \subst{X_2}{Y}, \subst{c}{Z}, \subst{d}{U}   \}}
                   {.9}.
   \]
\end{Example}



In Fig.~\ref{fuzfafotgenrules}, we give a fuzzy version of the
generalization rules taking into account mismatches not only in
functors, but also in arities; \ie, number and/or order of
arguments. Unlike Sessa's unification, we now assume that we are not
only given a similarity relation
$\sim\;\in\,\Sigma\times\Sigma\arrow[0,1]$ on the signature
$\Sigma=\cup_{n\geq 0}\Sigma_n$, but also that functors of different
arities may be similar with some non-zero truth value as specified by an
one-to-one argument-position mapping for each pair of so-similar
functors associating to each argument position of the functor of least
arity a distinct argument position of the functor of larger arity.  The
only rule among those of Figure~\ref{fuzfotgenrules} that differs is the
last one (\RuleName{Similar Functors}) which is now a pair of rules
called \RuleName{Functor/Arity Similarity Left} and
\RuleName{Functor/Arity Similarity Right} to account for similar
functors's argument positions depending which side has less
arguments. If the arities are the same, the two rules are equivalent.

\begin{figure}
\scriptsize
\vspace{-.4cm}
\begin{Rules}
\hspace{-.1cm}%
\GenRule[f\sim^p_\beta g; \;\; 0\leq m\leq n;\;\; \alpha_0\eqd\alpha\wedge\beta]
      {Functor/Arity Similarity Left}
      {\!\!\FFotGenUnapplyJudgement{\sigma_1}{\sigma_2}{\alpha_0}{s_1}{t_{p(1)}}{u_1}{\sigma^1_1}{\sigma^1_2}{\alpha_1}
       \,\ldots\,
       \FFotGenUnapplyJudgement{\sigma^{m-1}_1}{\sigma^{m-1}_2}{\alpha_{m-1}\!\!}{s_m}{t_{p(m)}}{u_m}{\sigma^m_1}{\sigma^m_2}{\alpha_m\!\!}}
      {\!\!\FFotGenJudgement{\sigma_1}{\sigma_2}
                        {\alpha}
                        {f(s_1,\ldots,s_m)}{g(t_1,\ldots,t_n)}
                        {f(u_1,\ldots,u_m)}
                        {\sigma^m_1}{\sigma^m_2}
                        {\alpha_m\!\!}}
\\ \\ 
\hspace{-.1cm}%
\GenRule%
[g\sim^p_\beta f; \;\; 0\leq n\leq m;\;\; \alpha_0\eqd\alpha\wedge\beta]
      {Functor/Arity Similarity Right}
      {\!\!\FFotGenUnapplyJudgement{\sigma_1}{\sigma_2}{\alpha_0}{s_{p(1)}}{t_1}{u_1}{\sigma^1_1}{\sigma^1_2}{\alpha_1}
       \,\ldots\,
       \FFotGenUnapplyJudgement{\sigma^{n-1}_1}{\sigma^{n-1}_2}{\alpha_{n-1}\!\!}{s_{p(n)}}{t_n}{u_n}{\sigma^n_1}{\sigma^n_2}{\alpha_n\!\!}}
      {\!\!\FFotGenJudgement{\sigma_1}{\sigma_2}
                        {\alpha}
                        {f(s_1,\ldots,s_m)}{g(t_1,\ldots,t_n)}
                        {g(u_1,\ldots,u_n)}
                        {\sigma^n_1}{\sigma^n_2}
                        {\alpha_n\!\!}}
\vspace{-.2cm}
\end{Rules}
\caption{\textbf{Functor/arity-weak generalization axioms and rule}}
\label{fuzfafotgenrules}
\end{figure}

\vspace{-.7cm}
\begin{theorem}\label{theorem4}
The fuzzy generalization rules of Fig.~\ref{fuzfotgenrules} where Rule
``\RuleName{Similar Functors}'' is replaced with the rules in
Fig.~\ref{fuzfafotgenrules} are correct.
\end{theorem}


\Section{Conclusion}

We have summarized the principal results regarding the derivation of
fuzzy lattice operations for the data structure known as first-order
term. This is achieved by means of syntax-driven constraint
normalization rules for both unification and generalization. These
operations are then extended to enable arbitrary mismatch between
similar terms whether functor-based, arity-based (number and order), or
combinations. The resulting lattice operations are in the same class of
complexity as their crisp versions, of which they are conservative
extensions---namely that of Union/Find. All these details, along with
proofs and examples, are to be found
in~\Cite{hakpasi:fuzfotlat:2017}{http://hassan-ait-kaci.net/pdf/fuzfotlat-preprint.pdf}.

As for future work, there are several avenues to explore. The most
immediate concerns implementation of such operations in the form of
public libraries to complement extant tools for first-order terms and
substitutions~\Cite{bousi:wam:2009}{https://www.researchgate.net/publication/221582279_A_Similarity-Based_WAM_for_BousiProlog}. This
is eased by the fact that the fuzzy lattice operations do no require
altering these conventional first-order structures. There are several
other disciplines where this technology has potential for fuzzifying
applications wherever {\fots} are used for their lattice-theoretic
properties such as linguistics and learning. Finally, most promising is
using this work's approach to more generic and more expressive knowledge
structures for applications such as Fuzzy Information
Retrieval~\Cite{fuzzyIR:2005}{https://www.irit.fr/publis/ADRIA/BougPetal001a.pdf}.
We are currently developing the same formal construction for fuzzy
lattice operations over order-sorted feature (\osf)
graphs~\Cite{osftu}{http://www.hassan-ait-kaci.net/pdf/osf-theory-unification.pdf}.
Encouraging initial results are being reported in~\cite{hak:gp:2017}.

\bibliographystyle{splncs03}
{\small\bibliography{main}}
\end{document}